\documentclass[aip, jcp, reprint]{revtex4-1}
\usepackage{graphicx}
\begin{document}
\title{Evaluating force field accuracy
with long-time simulations of a
\\tryptophan zipper peptide}
\affiliation{ 
Department of Physics, University of California, 
Davis, California, 95616, USA
}
\author{N. R. Hayre}
 \email{nrhayre@ucdavis.edu.}
\author{R. R. P. Singh}
\author{D. L. Cox}
\date{June 27, 2010}
\begin{abstract}
We have combined a custom implementation of the fast multiple-time-stepping
LN integrator with parallel tempering to explore folding properties
of small peptides in implicit solvent on the time scale of microseconds.
We applied this algorithm to the synthetic $\beta$-hairpin trpzip2 and one of its sequence variants W2W9.
Each simulation consisted of over 12 $\mu$s of aggregated virtual time.
Several measures of folding behavior showed convergence, allowing comparison with experimental equilibrium properties.
Our simulations suggest that the electrostatic interaction of tryptophan sidechains is responsible for much of the stability of the native fold.
We conclude that the ff99 force field combined with ff96 $\phi$ and $\psi$ dihedral energies and implicit solvent can reproduce plausible folding behavior in both trpzip2 and W2W9. 
\end{abstract}
\maketitle
\section{Introduction}
The study of small peptides with well-defined secondary-structural motifs gives insight into the basic properties of protein systems. 
The $\beta$-hairpin is one such motif that is of particular interest, consisting of two short anti-parallel $\beta$-strands connected by a turn.
In and of themselves, isolated $\beta$-hairpin sequences possess folding behavior similar to that of larger proteins, based on statistical mechanical considerations \cite{Munoz1997,Munoz1998}.
As a part of larger proteins, the $\beta$-hairpin motif may be important as a nucleation site along kinetic folding pathways \cite{Du2004}. 
Moreover, the relatively short timescale of folding of some $\beta$-hairpins \cite{Snow2004} coupled with their small size makes them amenable to all-atom simulation, lending insight into the physical system while providing a test of current models.  
Here we focus on a particular hairpin that has been subject to several experimental and theoretical studies.

The tryptophan zippers are a set of six 12- and 16-residue engineered $\beta$-hairpin peptides (trpzip 1 to 6), each possessing one or more cross-strand tryptophan pairs \cite{Cochran2001}.
The interaction of the indole side chains of tryptophan (Trp) imparts a stability that is unusual for such small proteins.
At least in the case of trpzip1, Trp pairings appear to be not only sufficient for stability, but necessary; when they are removed (via mutations of Trp to other hydrophobic residues), there is no evidence for hairpin formation \cite{Cochran2001}.
In analogy with mutations of trpzip1, a recent experiment found that a mutational variant of trpzip2 without potential cross-strand Trp pairings, dubbed W2W9 in reference to the residue indices where Trp residues were not mutated to valine, has no detectable folding, whereas mutants with intact cross-strand pairs are stable hairpins \cite{Wu2009}.

Their unusual stability, small size, and the availability of well-constrained native structural models has made the trpzip peptides the subject of numerous experimental and theoretical-computational studies.
Among the latter, several have demonstrated stability and folding of trpzip models using OPLS-AA \cite{Snow2004,Xu2008}, AMBER ff96 \cite{Yang2004,Zhang2006,Chen2008,Shell2008,YangShaoGaoQin2009}, and AMBER ff99SB \cite{Nymeyer2009} all-atom molecular mechanics force fields.
Despite its success in this context, ff96 may artificially stabilize extended $\beta$-strand conformations \cite{Ono2000,Mu2003,Gnanakaran2003}, and specifically $\beta$-hairpins \cite{GarciaSanbonmatsu2002}.

Herein, we demonstrate that the ff96 parameterization of backbone dihedral energies (denoted here dih96) can reproduce realistic equilibrium folding properties in trpzip2 and its W2W9 variant in silico, in agreement with experiment.
Simulation of the W2W9 variant embodies a novel test of force field accuracy, and gives further insight into $\beta$-hairpin stability.
Specifically, we find that the electrostatic interaction of cross-strand Trp pairs is critical for folding stability.

To carry out this study, we use a new implementation of the fast, 
accurate multiple-time-stepping (MTS) integration algorithm
LN (Langevin/Normal-mode) \cite{Barth1998}, coupled with replica exchange
molecular dynamics (REMD) for enhanced sampling.
REMD gives a picture of the thermal equilibrium properties of the system,
while the MTS integrator facilitates access to longer 
simulated times and thus more nearly-equilibrated structural ensembles.
Simulations were long enough to observe convergence of order parameters
from two simulations begun with very different initial conditions.
REMD has the disadvantage of obscuring realistic constant-temperature
folding dynamics and kinetic rates, so we have limited ourselves
to equilibrium properties.

\section{Methods}
\subsection{Energetics}
The 12-residue synthetic peptide trpzip2 (sequence S\-W\-T\-W\-E\-N\-G\-K\-W\-T\-W\-K\--NH2) and a mutated form W2W9 (sequence S\-W\-T\-V\-E\-N\-G\-K\-W\-T\-V\-K\--NH2) were modeled with classical molecular dynamics (MD).
Solute atoms interact by a potential energy whose functional form was described by \citet{Cornell1994}.
The parameters in this potential energy were drawn from the Assisted Model Building with Energy Refinement (AMBER) ff99 set \cite{Wang2000}.
Modifications to these parameters were considered for the potential energies of backbone $\phi$ and $\psi$ dihedral angle rotations.
The first modification, denoted ff99SB, is a proposed correction of the backbone dihedral energies based on quantum-mechanical calculations \cite{ISI:000241247100017}, establishing good relative energies between different secondary structures.
The second, denoted in this work as ff99/dih96, consists of corrections made by \citet{Kollman1997} (ff96) applied to the ff99 force field.
Specifically, no backbone dihedral energy terms are dropped from ff99; they are only modified where they differ from those in ff96.

The energy of electrostatic solvent interaction was approximated with the generalized Born (GB) model of \citet{Hawkins1996}.
A 0.10 M concentration of mobile counterions was assumed for calculation of Debye-H\"uckel screening.
The solute dielectric constant was set to unity, and that of the solvent was set to 78.5.

The solvent-accessible surface area (SASA) model is intended to account for the non-polar aspect of the conformationally-dependent free energy change of aqueous solvation \cite{Still1990}, and it is often used in conjunction with GB models under the name GBSA.
The SASA model was implemented here using the LCPO algorithm \cite{Weiser1999}, with a surface tension of 0.005 kcal/\AA$^2$.

\subsection{Dynamics -- Multiple Time Stepping}
The positions and velocities of atomic point masses were evolved with a Langevin equation of motion.
A multiple-time-stepping scheme, the LN (Langevin/normal-mode) algorithm \cite{Barth1998}, was used to accurately integrate the Langevin equation while improving computational efficiency.
LN improves upon the Langevin/implicit-Euler/normal-mode (LIN) scheme \cite{Zhang1994} by removing its computationally-expensive implicit integration.
LN achieves significant performance benefits over other MTS schemes by favoring force extrapolation to impulse, though this choice results in an integrator that is neither symplectic nor time-reversible, and which requires moderate damping to remain stable.

The present implementation of LN closely follows the introductory work \cite{Barth1998} in terms of the composition of force-splitting groups and distance cutoffs.
Following their proposal, bonded forces were directly calculated instead of linearized, for simplicity.
What follows is a brief description of our implementation of LN.

Terms in the total potential energy function $U$ are grouped according to the inherent length and time scales on which significant variations occur:
\begin{equation}
U = U_f + U_m + U_s + U_c,
\end{equation}
where the subscripts refer to ``fast'', ``medium'', ``slow'', and ``cutoff'' groups.
The fast group contains all bonded (pairwise, angle, and dihedral) terms.
The medium group contains all non-bonded (electrostatic, solvation, and van der Waals) terms wherein pairwise distance $d$ is less than an inner cutoff $d_{\mathrm{in}} = 5$\ \AA.
The slow group contains non-bonded terms for which $ d_{\mathrm{in}} \leq d < d_{\mathrm{out}} = 15$\ \AA, where $d_{\mathrm{out}}$ is an outer cutoff.
All energy terms for which $d \geq d_{\mathrm{out}}$ are placed in the cutoff group, and are not evaluated or used to calculate forces.
While bonded terms are kept permanently in the fast group, non-bonded terms were periodically reassigned among the other groups in order to satisfy the above pairwise distance conditions.

The total force $F$ on an atom was considered as the sum of contributions from each of the potential energy groups, excepting $U_c$:
\begin{equation}
F = F_f(r) + F_m(r_m) + F_s(r),
\end{equation}
where $F_i \equiv -\nabla U_i$.
The fast forces $F_f$ are evaluated at every timestep using instantaneous positions.
The medium forces $F_m$ are updated after every interval $\Delta t_m = k_m \Delta \tau$ using the positions $r_m  \equiv  r + \frac{1}{2} \Delta t_m v$ (midpoint extrapolation).
The slow forces $F_s$ are updated after every interval $\Delta t = k_s \Delta t_m$ using instantaneous positions (constant extrapolation).
Assuming an arbitrary integration timestep labeled $n$, atomic positions $r$ and velocities $v$ are updated according to:
\begin{eqnarray}
r_{n+1/2} & = & r_n + \frac{\Delta \tau}{2} v_n \nonumber \\
F_f & = & F_f(r_{n+1/2}) \nonumber \\
F_m & = & F_m(r_n + \frac{1}{2} \Delta t_m v) \ \mathrm{if} \ n \bmod k_m = 0 \nonumber \\
F_s & = & F_s(r_{n+1/2}) \ \mathrm{if} \ n \bmod k_m k_s = 0 \nonumber \\
v_{n+1} & = & \frac{v_n + \Delta \tau (F_f + F_m + F_s + R_n) / m}{1 + \gamma \Delta \tau} \nonumber \\
r_{n+1} & = & r_{n+1/2} + \frac{\Delta \tau}{2} v_n \nonumber \\
(r_{n+1}, v_{n+1}) & \gets & \mathrm{RATTLE}(r_{n+1}, v_{n+1}). \nonumber \\
\end{eqnarray}
$R_n$ is a zero-mean normally-distributed random process with variance $ 2 k_B T m \gamma / \Delta \tau  $, where $\gamma = 20$ ps$^{-1}$ is a collision frequency that determines the strength of temperature coupling.
Pairwise terms were reassigned to appropriate groups every $k_m \cdot k_s$ timesteps; and Born radii and solvent pairlists were recalculated after each evaluation of the medium group forces. 

Positions and velocities of hydrogen atoms and their heavy-atom bonding partners were constrained using the RATTLE algorithm \cite{Andersen1983}, which allowed an augmented base timestep of $\Delta \tau = 2 \ \mathrm{fs}$.

The above algorithm was implemented in-house entirely in the C++ language and built with the GNU Compiler Collection, version 4.3.1.
Source code is available upon request.
Both the validation of this MD algorithm (as described in the Appendix) and computational speed measurements were performed using the AMBER 9 package \cite{AMBER9} as a standard.

\subsection{Performance}
\begin{figure}
\centering
\includegraphics{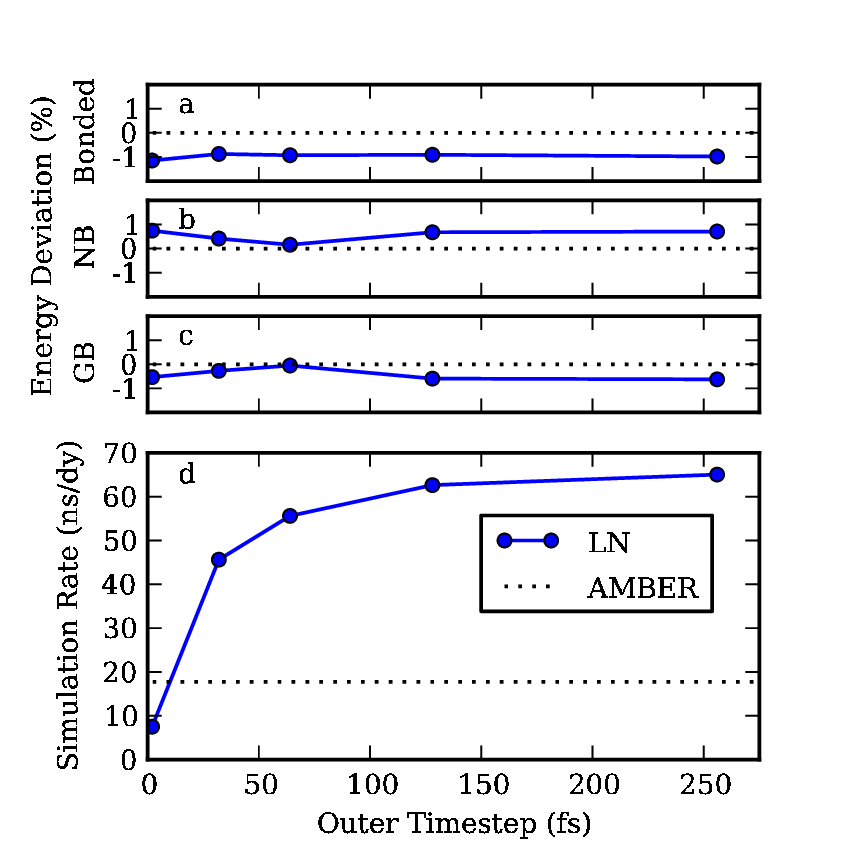}
\caption{
The upper three graphs show the small deviations of LN energies from AMBER energies by type, expressed as percent errors from the AMBER values. 
(a) Bonded energy is the sum of pairwise, angle, and dihedral terms;
(b) nonbonded energy (denoted NB) is the sum of vacuum van der Waals and electrostatic terms;
and (c) GB represents the energy of generalized Born solvation.
The lower graph (d) shows a rapid increase of computational rate with the outer timestep of LN.
}
\label{fig:AMBER_vs_LN_direct}
\end{figure}

The LN algorithm was compared to AMBER in terms of computational speed using a single core of an AMD Opteron 2216 system running at 2.4 GHz.
As shown in Figure \ref{fig:AMBER_vs_LN_direct}, AMBER is faster than LN by about a factor of two at the lowest outer timestep (corresponding to no constant extrapolation and equal timesteps), presumably due to more optimized computer code.
With increasing outer timestep, however, LN rapidly surpasses AMBER and approaches an asymptotic maximum above 60 ns/day.
A medium timestep $\Delta t_m = 4$ fs ($k_m = 2$) was chosen to balance stability and numerical accuracy with efficiency. 
An outer timestep of $\Delta t = 128 \ \mathrm{fs}$, corresponding to $k_s = 32$, was settled upon to balance efficiency gain and conservative extrapolation.

\subsection{Replica Exchange}
A replica exchange scheme was used to improve sampling
and observe temperature-dependent effects \cite{Sugita1999}.
Replicas were simulated at 20 target temperatures $T_i$ with $i = 1, 2, \dots, 20$.
$T_1$ and $T_{20}$ were set to 200 K and 500 K, respectively, and the intermediate values were separated with exponentially increasing intervals.
During simulation, pairs of replicas at adjacent temperatures $T_i$ and $T_{i+1}$ were exchanged with a probability $p = \min(1, \exp[ (\beta_i - \beta_{i+1}) (U_i - U_{i+1}) ])$.
The values $\beta_i$ and $U_i$ are the inverse temperature and total potential energy, respectively, of the replica at $T_i$.
A round consisted of attempting an exchange every 5 ps between temperature-adjacent replica pairs in order of increasing temperature, except for the last attempt, which was between the highest- and lowest-temperature replicas.
Rounds of exchange attempts were carried out continuously.

The convergence of order parameters computed from two independent REMD simulations in which the initial replica conformations are significantly different (as measured by the same order parameter) can be used to indicate the degree to which equilibrium is sampled \cite{Juraszek2009}.
REMD runs were thus performed in pairs, referred to as ``unfolding" and ``folding" runs.
The unfolding run started with all replicas in the native
$\alpha$-carbon (C$_{\alpha}$) conformation of trpzip2 
(PDB ID: 1LE1, model 1 \cite{Cochran2001}).
The folding set started in a fully extended conformation in which all the peptide backbone dihedral angles are set at $\pi$ radians.
Initial steric clashes in both ensembles were reduced with a steepest-descent minimization of the total potential energy.

\subsection{Analysis}
Configurations and energies were recorded for each replica at 10 ps intervals, and these data were used to calculate several quantities.
The C$_{\alpha}$ radius of gyration $R_g$ is a simple measure of the spatial compactness of a conformation.
Given $N$ C$_{\alpha}$ positions ${\vec{r}_i}$, and defining their pairwise distance as $d_{ij} = |\vec{r}_j - \vec{r}_i|$, $R_g$ is calculated as
\begin{equation}
R_g^2 = \frac{1}{2N^2} \sum_{i=1}^N \sum_{j=1}^N d_{ij}^2.
\end{equation}

A specific measure of the difference of a given conformation from the native is provided by the average pairwise C$_{\alpha}$ distance root-mean-square deviation $D_{\mathrm{rms}}$, calculated as
\begin{equation}
D_{\mathrm{rms}}^2 = \frac{1}{2N^2} \sum_{i=1}^N \sum_{j=1}^N (d_{ij}^{(m)} - d_{ij}^{(n)})^2
\end{equation}
where the lettered superscripts label pairwise distances calculated on two different conformations, one of which is native.
This is a variation on a previous formula \cite{ISI:000178737200015}, slightly modified in its prefactor to be more analogous to $R_g$.

There are five hydrogen bonds present between the two antiparallel $\beta$-strands in the native structure, as well as a terminal salt bridge. 
To provide a sensitive measure of the formation of these bonds, we define the deviation of each as the difference of the distance between the donor and acceptor atoms and the force field equilibrium distance.
We further define the fraction of formed native H-bonds $f_{\mathrm{H}}$ as the average of the count of these bonds, including the terminal salt bridge, whose deviations are less that 0.3 \AA.

For the results involving ff99SB presented below, we use a measure of alpha-helical contacts to characterize alternate conformations that are markedly different from the native structure.
A contact is counted as formed if the pairwise distance of 
any carbonyl oxygen to any amide nitrogen located four residues closer to the C-terminus is less than 3.3 \AA.
This measure is limited, since we do not employ an angular criterion to discriminate the relative orientation of the donor, proton, and acceptor.

We employ two simple indicators of the Trp sidechain properties.
First, the distances between $C_{\delta 1}$ sidechain atoms in each of the cross-strand Trp-Trp residue pairs (4,9) and (2,11) indicate how closely-packed the side chains are.
Second, the relative orientation of the planes of the indole rings in each pair is indicated by the coplanarity, or the normalized dot product of two vectors that lie in each plane, each pointing from $C_{\gamma}$ to $C_{\delta 2}$ in its respective sidechain.

All potentials of mean force are computed using an existing implementation of the multistate Bennett acceptance ratio (MBAR) method \cite{Shirts2008}.

\section{Results}
Initial REMD on trpzip2 with ff99SB and a comparison of various AMBER
force fields with constant-temperature MD demonstrate the sensitivity
of the molecular mechanics model to changes in $\phi$ and $\psi$
dihedral energy.
REMD on trpzip2 with ff99/dih96 shows qualitatively accurate folding behavior
that depends weakly on the nonpolar component of solvation energy.
REMD on W2W9 with ff99/dih96 gives results that are consistent with experiment
and reveals factors involved in stabilizing the native state.

\subsection{trpzip2 with ff99SB}
\begin{figure}
\centering
\includegraphics{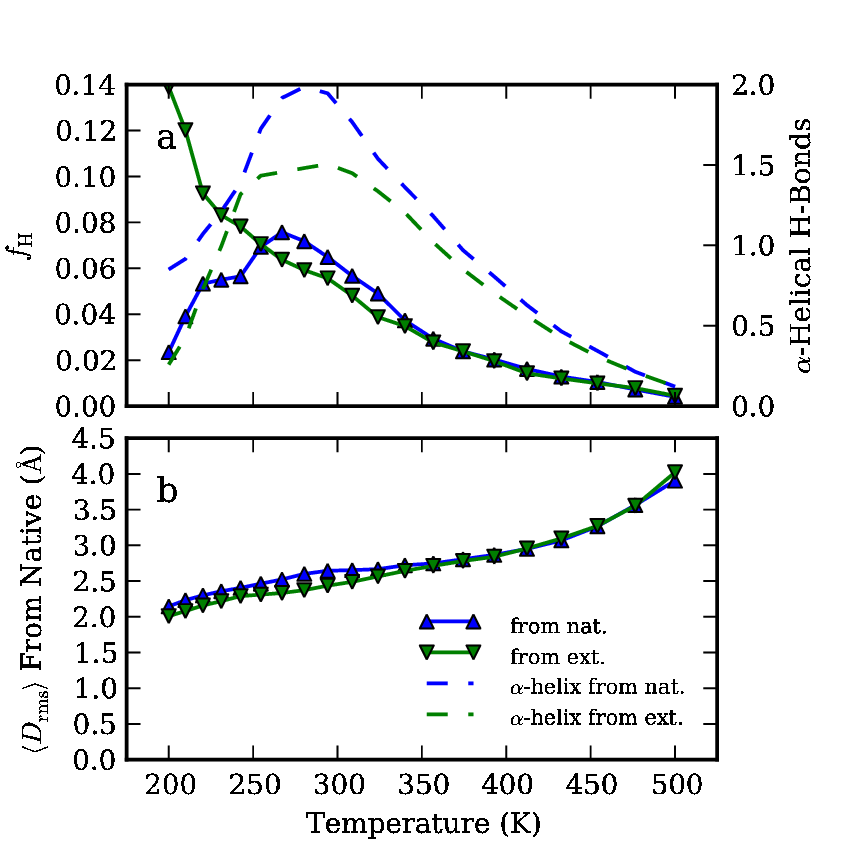}
\caption{
Thermal variation of order parameters using the ff99SB/GB force field.
(a) The fraction of formed native H-bonds $f_{\mathrm{H}}$ in trpzip2 is small, while the number of formed $\alpha$-helical contacts is appreciable.
(b) The $D_{\mathrm{rms}}$ from the native structure indicates a markedly non-native structure at all temperatures.
All ordinates are averages over the final 200 ns per replica of simulated time.
Note that helical contacts are not well-converged, which may be due to a slower timescale for equilibration or slow fluctuation in this order parameter.
}
\label{fig:SB_obsv_vs_T}
\end{figure}

Folding and unfolding REMD runs of length 700 ns were first performed using ff99SB/GB.
Figure \ref{fig:SB_obsv_vs_T} shows the native $D_{\mathrm{rms}}$ and formed native backbone hydrogen bonds $f_{\mathrm{H}}$ for both initial conditions, averaged over the final 200 ns of simulated time.
These observables converge on values that indicate non-native equilibrium states at all temperatures.
Average alpha-helical contacts in trpzip2 are also plotted versus temperature, showing that the competing stable state at standard temperature is partially alpha-helical.

\begin{figure}
\centering
\includegraphics{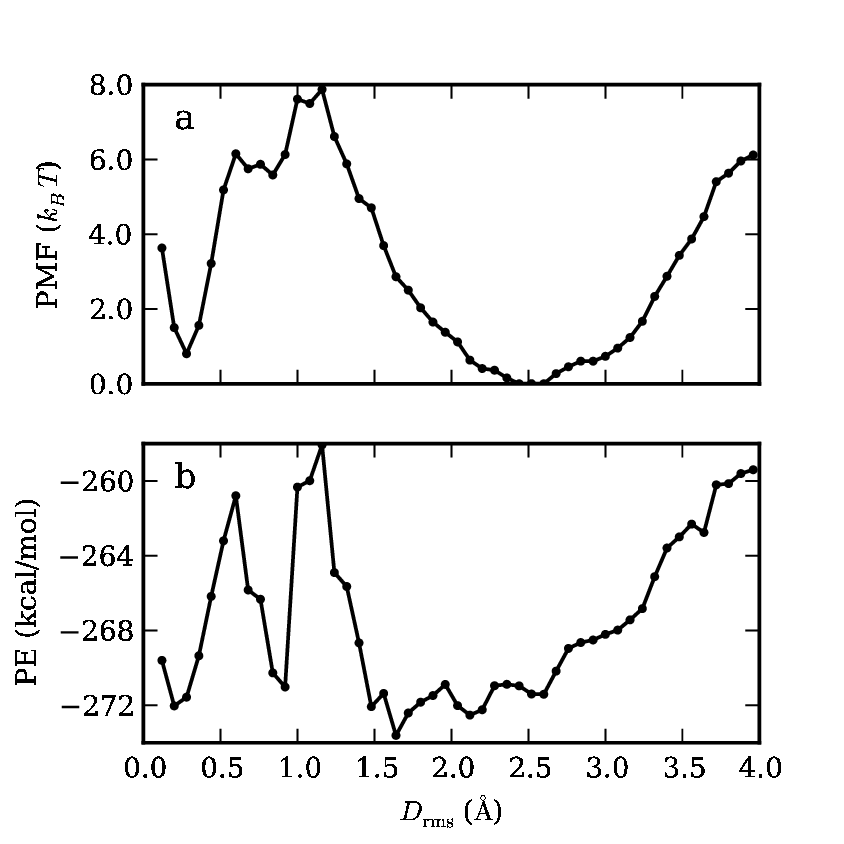}
\caption{
The potential of mean force versus $D_{\mathrm{rms}}$ at 293 K shown in (a)
indicates that the native state is not globally stable under
the ff99SB force field in GB solvent.
The canonical expectation for total potential energy (PE) shown in (b)
is compared in the text with previous results with the same force field.
These values were generated from 28 $\mu$s of aggregated data from both folding and unfolding runs.
}
\label{fig:SB_PMF}
\end{figure}

Figure \ref{fig:SB_PMF} shows a potential of mean force and average total potential energy versus the $D_{\mathrm{rms}}$ observable.
The PMF gives evidence for a structural state that is more stable than the native and quite dissimilar from it.
The potential energy profile shown may be compared to a similar result derived from decoy analysis of trpzip2 under the same force field and the same implicit solvent model \cite{ISI:000241247100017}.
In that analysis, a different metric of deviation from the native conformation (rmsd) was used, and the potential energies do not account for thermal effects, as noted therein.
Nevertheless, a small energetic stability of the native state was suggested, whereas the current result indicates marginal instability of the native state in both potential energy and free energy under canonical simulation conditions.

\subsection{Comparison of AMBER force fields}
We compared several AMBER force fields by computing the difference in total potential energy between the native structure and a competitively stable alternate structure in short, constant-temperature simulations.
The alternate structure was chosen as the final conformation of the replica with the highest occupancy at the lowest temperature over the final 200 ns of the REMD folding simulation employing ff99SB.
Trajectories were generated for both the native and alternate conformations with ff99, ff99SB, ff99/dih96, and ff03 force fields, and average energy differences were calculated, as shown in Table \ref{tab:compare}.
In addition, a set of runs was performed with each force field using the SASA approximation for non-polar solvation energy.
This comparison of parameter sets indicates a large stability of the native conformer with ff99/dih96.
It is also evident that SASA energy does not greatly lower the energy of the native conformer for any of the tested force fields.
\begin{table*}
\caption{\label{tab:compare}
The energy difference between the native and the chosen alternate conformer of trpzip2 with variants of the AMBER force field and generalized Born implicit solvent.
---/SA indicates that the SASA model was included in the simulation.
$\Delta E = E_{\mathrm{nat}} - E_{\mathrm{alt}}$, so that a negative value indicates a more stable native conformer.
For each conformer, averages and errors are calculated on 32 simulations of duration 100 ps at temperature 293 K, using the AMBER 9 package.
}
\begin{ruledtabular}
\begin{tabular}{l|cccccccccc}
& ff99 & ff99/SA & ff99SB & ff99SB/SA 
& ff99/dih96 & ff99/dih96/SA & ff96 & ff96/SA
& ff03 & ff03/SA \\
\hline
$\Delta E$ (kcal/mol) 
\footnote{
For $\Delta E$, the average error of the mean is $0.35$ kcal/mol.
}
& $+11.6$ & $+11.9$ 
& $-3.5$ & $-4.3$ 
& $-28.1$ & $-28.6$ 
& $-19.2$ & $-19.6$
& $-1.7$ & $-1.6$ \\
$\Delta E_{SA}$ (kcal/mol) 
\footnote{
For $\Delta E_{SA}$, the average error of the mean is $0.02$ kcal/mol.
}
&  & $-0.22$ 
&  & $-0.32$ 
&  & $-0.35$ 
&  & $-0.33$ 
&  & $-0.33$ \\
\end{tabular}
\end{ruledtabular}
\end{table*}

\subsection{trpzip2 with ff99/dih96}
\begin{figure}
\centering
\includegraphics{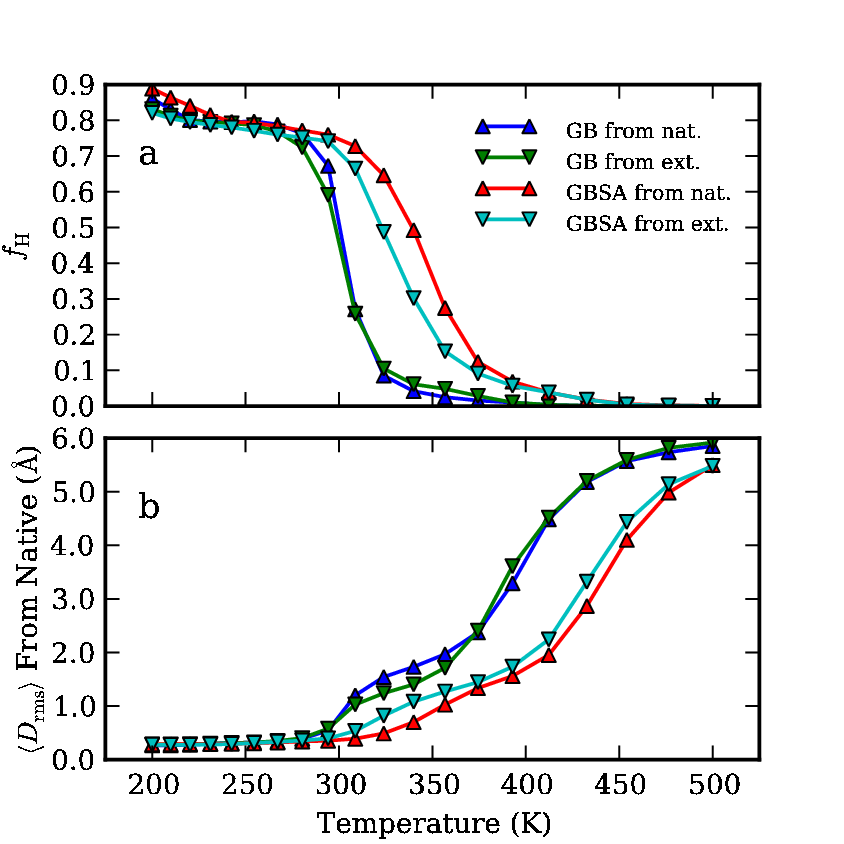}
\caption{
(a) The fraction of formed native H-bonds $f_{\mathrm{H}}$ in trpzip2
and (b) the $D_{\mathrm{rms}}$ from native versus temperature
for the ff99/dih96 parameter set with GB and GBSA solvent models.
Both order parameters clearly indicate one or more conformational transitions.
All ordinates are averages over the final 200 ns per replica of simulated time.
The near convergence of the averages of these coordinates demonstrates the degree of equilibration reached in this interval.
}
\label{fig:96_obsv_vs_T}
\end{figure}

REMD was performed again using ff99/dih96/GB and ff99/dih96/GBSA, given their prediction of larger relative native-state energetic stability in the analysis above.
Graphs of native contacts and $D_{\mathrm{rms}}$ versus temperature in Figure \ref{fig:96_obsv_vs_T} are indicative of a folding transition favoring the native conformation.
The disparity of midpoint temperatures in the thermal transitions for both solvent models suggests that folding involves more than two states.

\begin{figure}
\centering
\includegraphics{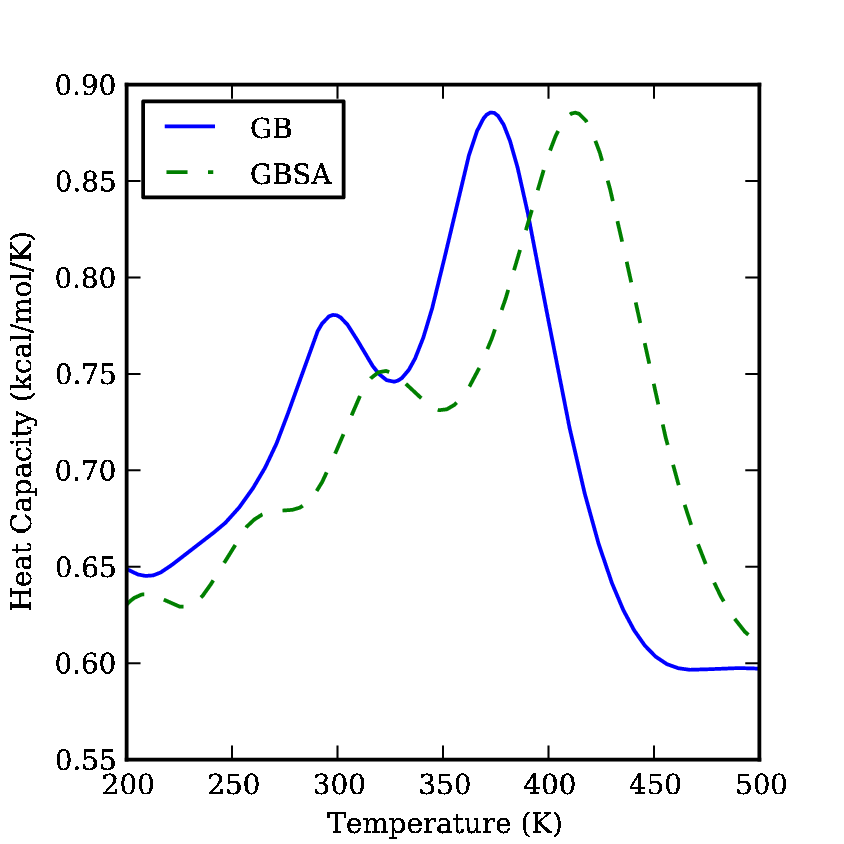}
\caption{
The heat capacity of ff99/dih96 vs. temperature with and without
SASA energy (GB and GBSA, respectively) shows two significant transition
temperatures.
Heat capacity is calculated as the first derivative
of the interpolating spline of the average
potential energy versus temperature, using the aggregated
final 200 ns per replica of folding and unfolding runs.
}
\label{fig:C}
\end{figure}

\begin{figure}
\centering
\includegraphics{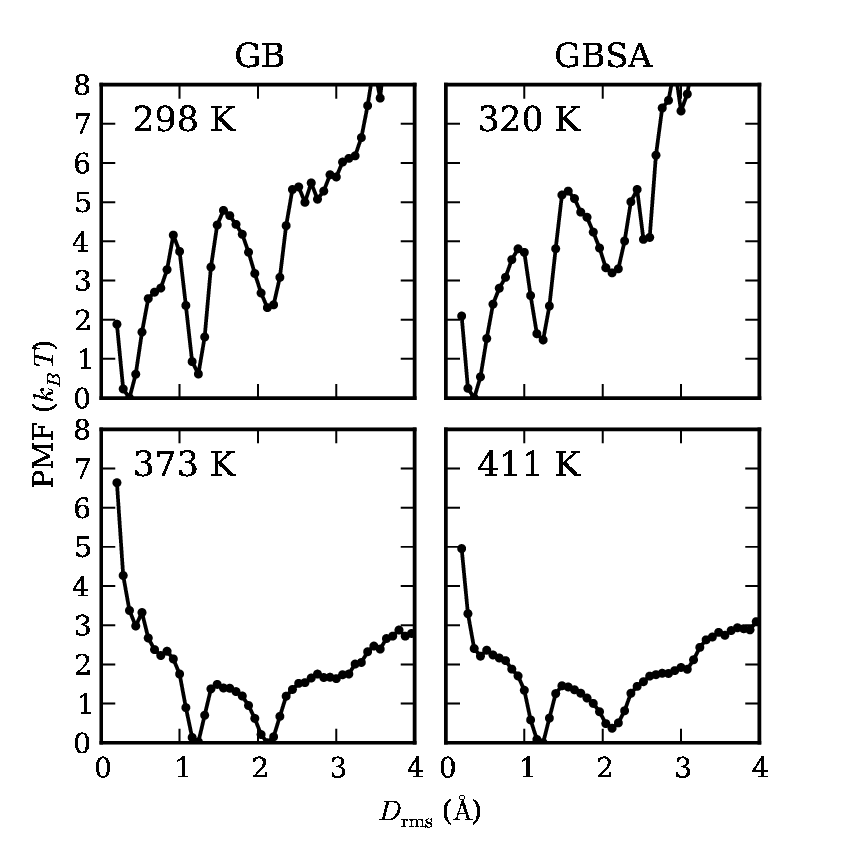}
\caption{
Potentials of mean force against $D_{\mathrm{rms}}$ are shown for several temperatures, with columns headers labeling the solvent model used.
Each potential is labeled by the temperature at which it was evaluated, which is close to a major peak in the heat capacity for the respective solvent model, as shown in Figure \ref{fig:C}, and corresponds to the onset of a shift of the global free energy minimum.
}
\label{1D_PMFs}
\end{figure}

Heat capacity versus temperature, shown in Figure \ref{fig:C}, reinforces this, suggesting two significant transitions occur -- near 300 K and 375 K for GB, and near 320 K and 410 K for GBSA.
These pairs of transition temperatures correspond well with the midpoints of the thermal transitions in $f_{\mathrm{H}}$ and $D_{\mathrm{rms}}$, respectively.
One-dimensional (1D) potentials of mean force along the $D_{\mathrm{rms}}$ coordinate, computed for temperatures near the heat capacity peaks, show the onset of transitions as exchanges of free energy minima (see Figure \ref{1D_PMFs}).
These potentials reveal the presence of an unfolded state and at least one intermediate state.

\begin{figure}
\centering
\includegraphics{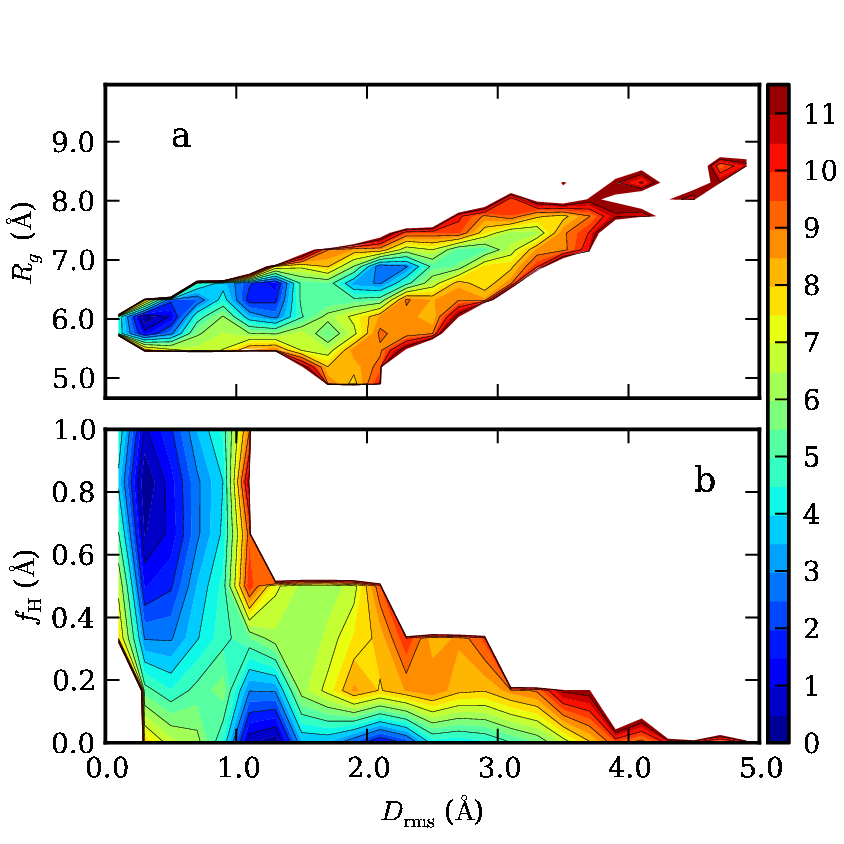}
\caption{
Two-dimensional potentials of mean force plotted against 
(a) $R_g$ and $D_{\mathrm{rms}}$; and (b) $f_{\mathrm{H}}$ and
$D_{\mathrm{rms}}$, indicating the abrupt formation of most native H-bond
contacts upon the transition from the intermediate to the native state.
Contours are placed at intervals of $k_B T$, and the temperature 
$T = 298\ \mathrm{K}$ corresponds to the first major peak in the heat capacity
for the GB solvent model.
}
\label{fig:2DPMFs}
\end{figure}

Further evidence for three states can be observed in the two-dimensional potentials of mean force shown in Figure \ref{fig:2DPMFs}.
A PMF versus $D_{\mathrm{rms}}$ and $R_g$ shows the native state, a significant folding intermediate, and an unfolded state that are progressively larger in spatial extent, commensurate with increasing $D_{\mathrm{rms}}$.
Projected onto $D_{\mathrm{rms}}$ and $f_{\mathrm{H}}$, the PMF shows that the transition from the intermediate to the native state involves the formation of most or all the native H-bonds, with no partially-bonded intermediates.

\begin{figure}
\centering
\includegraphics{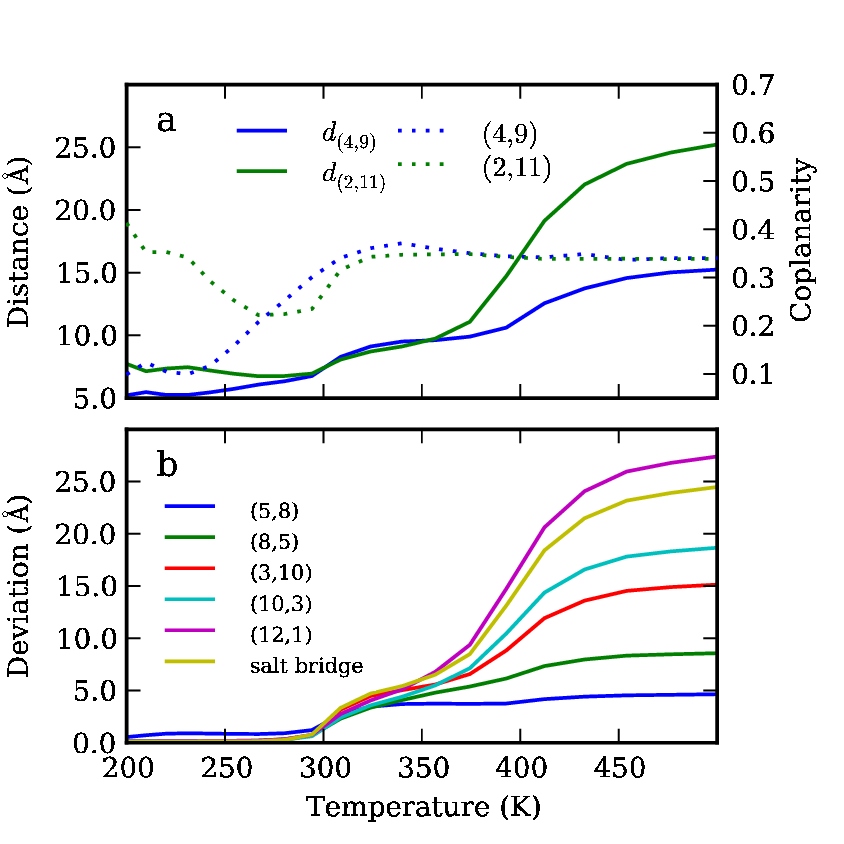}
\caption{
(a) Pairwise $C_{\delta 1}$ distances (solid lines) and coplanarity (dotted lines, secondary ordinate) of cross-strand Trp-Trp pairs and
(b) deviations of H-bond and salt bridge distances from equilibria
show more detailed structural features of the two major transitions.
}
\label{fig:more_obsv}
\end{figure}

The nature of the folded, intermediate, and unfolded states is revealed in greater detail by inspecting the thermal change of coordinates involving the native H-bond deviations $d_{\mathrm{H}i}$ and the two cross-strand Trp-Trp sidechain C$_{\delta}$ distances $d_{(4,9)}$ and $d_{(2,11)}$ (see Figure \ref{fig:more_obsv}).
Only the GB model is considered, though the properties are similar for GBSA.
Above 375 K, these coordinates suggest an unfolded state, where all $d_{\mathrm{H}i}$ and C$_{\delta 1}$ distances are large and trend with sequence separation, with smaller distances separating pairs closer to the $\beta$-turn.
Between 300 K and 375 K, both sets of coordinates are restricted to a narrow range of distances.
In this thermal regime, H-bond distances are uniform and more than double their equilibrium bonded lengths on average, and the two C$_{\delta 1}$ distances are also nearly equal.
This indicates that the hairpin is in a non-native collapsed state, wherein both hydrogen bonds and cross-strand Trp-Trp pairs reside close to their lower-energy folded states.
The uniformity and small values of both H-bond deviations and C$_{\delta 1}$ distances in this temperature range are consistent with a native-like, nearly-folded average structure for the intermediate.
Finally, below 300 K, H-bond distances settle close to their equilibria, and Trp-Trp distances take on values particular to their packing geometry.

\subsection{W2W9 with ff99/dih96}
\begin{figure}
\centering
\includegraphics{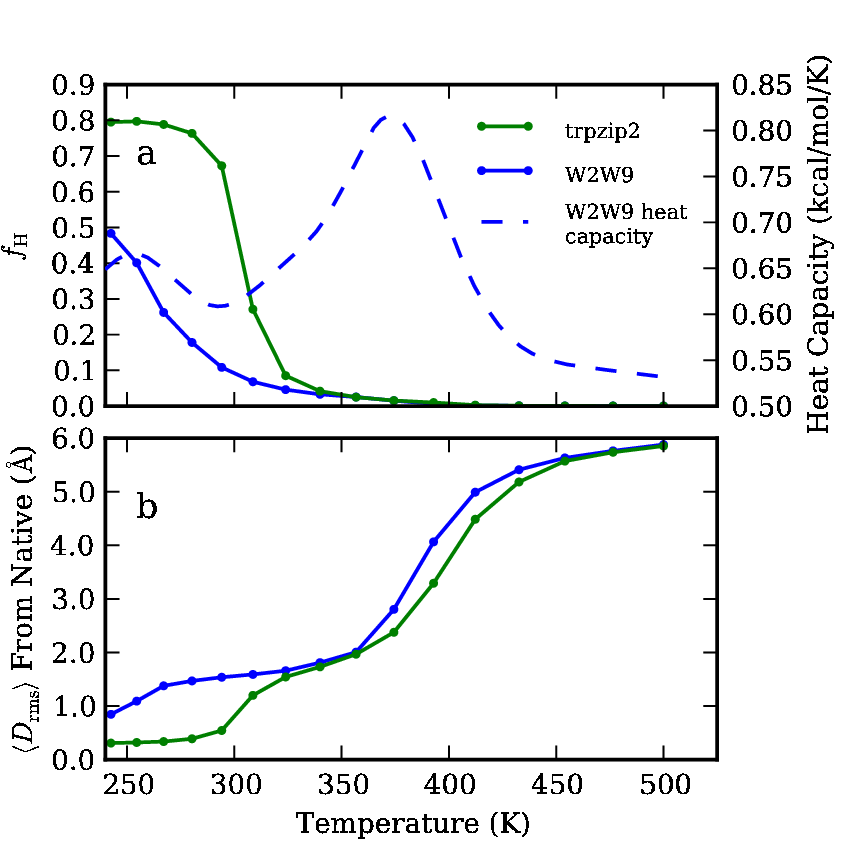}
\caption{
(a) $f_{\mathrm{H}}$, heat capacity, and (b) $D_{\mathrm{rms}}$ vs. temperature 
for W2W9, using ff99/dih96/GB.
Averages over folding and unfolding runs are shown, and trpzip2 data using the same model parameters are shown for comparison.
}
\label{fig:W2W9_obsv_vs_T}
\end{figure}

\begin{figure}
\centering
\includegraphics{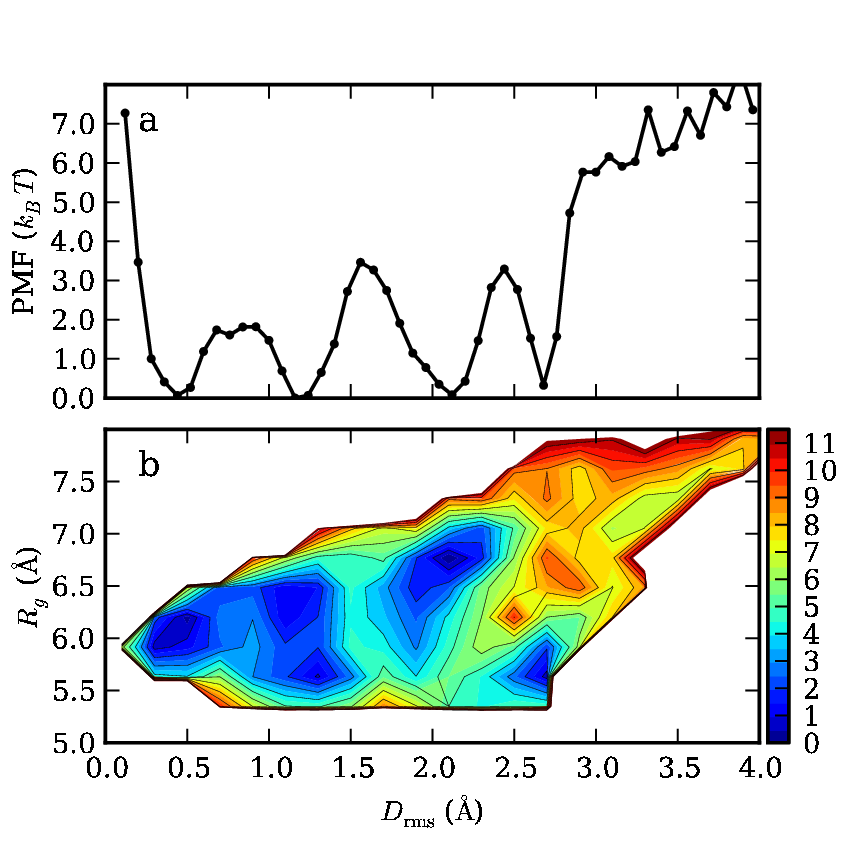}
\caption{
(a) The 1D PMF versus $D_{\mathrm{rms}}$; and 
(b) the 2D PMF versus $D_{\mathrm{rms}}$ and $R_g$ 
for W2W9 using ff99/dih96/GB, 
both computed at 275 K from 200 ns per replica from both folding and unfolding runs.
}
\label{fig:W2W9_PMF}
\end{figure}

The W2W9 mutated form of trpzip2 was simulated with REMD for 600 ns per replica using ff99/dih96/GB.
Figure \ref{fig:W2W9_obsv_vs_T} shows heat capacity, $f_{\mathrm{H}}$, and $D_{\mathrm{rms}}$ versus temperature, with the corresponding results for trpzip2 shown for comparison.
PMFs versus $D_{\mathrm{rms}}$ and $R_g$ in Figure \ref{fig:W2W9_PMF} reveals three basins that are, by the $D_{\mathrm{rms}}$ measure alone, indistinguishable from the unfolded, intermediate, and native states of trpzip2.
The 2D PMF shows that two more compact non-native states join the intermediate and unfolded free-energy basins in possessing competing stability relative to the native.
The native hairpin in W2W9 is clearly less stable, with a melting temperature near 250 K, though the 1D PMF gives an estimate of 31\% for the population of structures with $D_{\mathrm{rms}} < 1.0$ at 275 K. 

The loss of cross-strand Trp pairs uncovers the driving factors for stabilizing the intermediate and native states.
The high-temperature collapse to a native-like intermediate occurs near the same temperature as for trpzip2, indicating that dihedral energies drive this collapse.
The transition to a folded structure occurs much lower in temperature, and the intermediate state has a protracted range of thermal stability.
The Trp pairs are thus essential for stabilizing the native state and increasing the cooperativity of the lower-temperature transition.

\section{Discussion}

The foregoing results present several remarkable features.
In implicit solvent, and holding other conditions constant, we have observed a substantial sensitivity in folding behavior to variation of the $\phi$ and $\psi$ dihedral potential energies.
In particular, while ff99SB dihedral energies do not support a stable native state for trpzip2 in the given simulation conditions, those derived from ff96 do so quite well.
Through simulations of the W2W9 mutant, we have found these backbone parameters are not sufficient for hairpin stability; electrostatic sidechain interactions are also critical. 
Moreover, our results with W2W9 compare well with experiment, suggesting that ff99/dih96 is robust for modeling $\beta$-hairpin structures.

The ff99SB force field with GB implicit solvent results in a stable non-native $\alpha$-helical state.
This provides a potential contrast with a recent computational study of trpzip2 that used the same force field and TIP3P explicit solvent \cite{Nymeyer2009}, which found a folding free energy close to -0.5 kcal/mol at 273 K.
However, this divergence is unsurprising given that previous studies comparing implicit and explicit solvent have shown that GB models tend to overstabilize $\alpha$-helical conformations \cite{Zhou2003,Nymeyer2003,Roe2007}.
As indicated by the potential of mean force versus $D_{\mathrm{rms}}$ at 293 K in Figure \ref{fig:SB_PMF}, this effect need not be large to destabilize the native state in the context of ff99SB.
In fact, we observe a folding free energy close to +0.5 kcal/mol at 273 K (leading to an unstable native state) -- a discrepancy of close to 1.0 kcal/mol between GB implicit and TIP3P explicit solvent.
We have not tested the effect of a SASA model applied along with ff99SB; however, the results in Table \ref{tab:compare} show only a slight energetic stabilization of the native conformer with SASA energy.

SASA energy is evidently not a dominant factor in stabilizing the native state of trpzip2 in the context of the ff99/dih96 GBSA model used here.
SASA energy shifts transition temperatures higher and increases free energy changes between these states, but does not significantly change their location on the free energy landscape.
This is a sensible result of the surface area penalty imposed by the SASA model, which is expected to generically stabilize compact structures.
Unfortunately, until tested with an accurate force field, the underlying importance of nonpolar effects as modeled with SASA energy will remain obscure. 

Our simulations of the W2W9 variant of trpzip2 are in qualitative agreement with experiment.
W2W9 is not expected to have a significant folded population at any temperature down to 275 K, which is the lower limit of the experimentally investigated range of temperatures \cite{Wu2009}.
Indeed, in our REMD simulation of W2W9, we observe a diminished near-native population ($D_{\mathrm{rms}} < 1.0$) of about 31\% at 275 K using ff99/dih96/GB.

Our results on W2W9 suggest that the electrostatic interaction of Trp pairs greatly stabilizes the native hairpin structure, since the principal effect of the mutation from trpzip2 to W2W9 is the removal of such pairs.
The importance of a purely electrostatic effect is clear, since the SASA model was not employed in this simulation.
This finding adds greatly to the putative importance of this interaction, which has been implicated in the edge-to-face packing arrangement of the indole sidechain pairs \cite{Guvench2005}. 

It is worth considering the influence of the implicit solvent model on the stability of the W2W9 hairpin.
First, we have not included a model of non-polar effects, but SASA energy would likely further stabilize the hairpin, as we showed in the case of trpzip2.
Second, as mentioned earlier, GB models tend to shift secondary structure preference towards $\alpha$-helical conformations when compared with explicit solvent models.
This tendency would be expected to further destabilize the observed $\beta$-like native conformation.
The inaccuracy of the solvent model is thus probably underpredicting the stability and population of the folded hairpin.

Force fields employing the ff96 dihedral parameters have produced at least qualitatively accurate results for trpzip2 \cite{Yang2004,Shell2008,XiaoChenHe2009,YangShaoGaoQin2009}, and our results fall into agreement with these.
Furthermore, our study of W2W9 provides evidence for the robustness of these parameters in modeling $\beta$-hairpin structures.
We have presented a simple but critical test of force field accuracy in reproducing $\beta$-hairpin folding based on experimental findings.
This is only one of many possible comparisons with experiment in small peptide systems -- variation in primary sequence, pH conditions, and temperature, for example -- that can be useful for evaluating theoretical-computational approaches.
Moreover, the availability of increasingly powerful computational resources and techniques enables such evaluations based on equilibrium properties.

\begin{acknowledgments}
N. R. H. would like to acknowledge support received
during part of this project from
the Institute for Complex Adaptive Matter (ICAM-I2CAM) through the
ICAM Branches Cost Sharing Fund and the US National Science Foundation
ICAM Award, Grant DMR-0456669.
N. R. H. would also like to thank the cluster system administrators in the Computational Science and Engineering Department at U. C. Davis 
for technical support, and Dr. David Mobley for his many thoughtful comments.
\end{acknowledgments}

\appendix*

\section{Algorithm Validation}

\subsection{Comparison of Absolute Energies}
In order to validate the LN algorithm in the form described in this work, we compared it with the Langevin dynamics algorithm in the AMBER 9 package \cite{AMBER9} as a standard.
For identical simulation conditions (initial structure, force field, temperature, duration, etc.), we might expect the time-averaged energies obtained by using each MD algorithm to be the same (within error) over many independent trials.

To test this, we performed 32 independent 100-ps simulations for each algorithm, starting from model 1 of the experimental trpzip2 structure \cite{Cochran2001}.
The temperature was held fixed at 293 K, the ff99SB force field was used, and all other simulation parameters were set to the values mentioned in the Methods section.
For each algorithm, we computed the average of the bonded, nonbonded, and solvation energies over the duration of each simulation and then over all 32 simulations.
In Figure \ref{fig:AMBER_vs_LN_direct}a-c, each of these three average energies for the LN simulations is expressed as a percent deviation from the values for the AMBER simulations.
This test was repeated for several choices of the timesteps $\Delta t_m$ and $\Delta t$ in the LN algorithm.

There is a systematic energy shift between AMBER and LN, but it within two percent of the AMBER value for all energy components, including the total potential energy.
The mean energy in LN also varies with outer timestep, but only within half a percent of the base value at $\Delta t = 2 \ \mathrm{fs}$.
The average error in LN temperature when $\Delta t_m = 4$ fs is $+0.58\% \pm 0.01\%$.

\subsection{Comparison of Energy Differences}
Systematic deviations of internal energy and temperature in LN simulations with larger medium timesteps were duly noted in the
original work (Figure 6 in Ref.~\onlinecite{Barth1998}).
Such deviations are worth serious consideration, since stability in biomolecules can be based on energy differences as small as a few kcal/mol.
Accordingly, what is most crucial in the evaluation of LN is how well this algorithm preserves differences in internal energy between different structures, rather than how well it preserves individual energy values for a given structure.

To see how well LN preserves energy differences,
we obtained a diverse set of trpzip2 structures,
simulated them with both LN and AMBER according to
the protocol described in the previous section,
and calculated the differences in average total energies
between all the unique pairs in this set of structures.
Ideally, when LN was used, these differences
would be nearly the same as when AMBER was used.

To obtain the diverse set of trpzip2 structures, a pool of 160 snapshots was obtained at all temperatures at several well-separated times during the long REMD simulations performed in this study.
Clusters were formed from this pool using the $D_{\mathrm{rms}}$ metric according to an algorithm described by Daura \cite{Daura1999}, with a cutoff of 1.5 \AA.
The 12 cluster centers that resulted from this analysis formed the diverse structure set. 

\begin{figure*}
\centering
\includegraphics{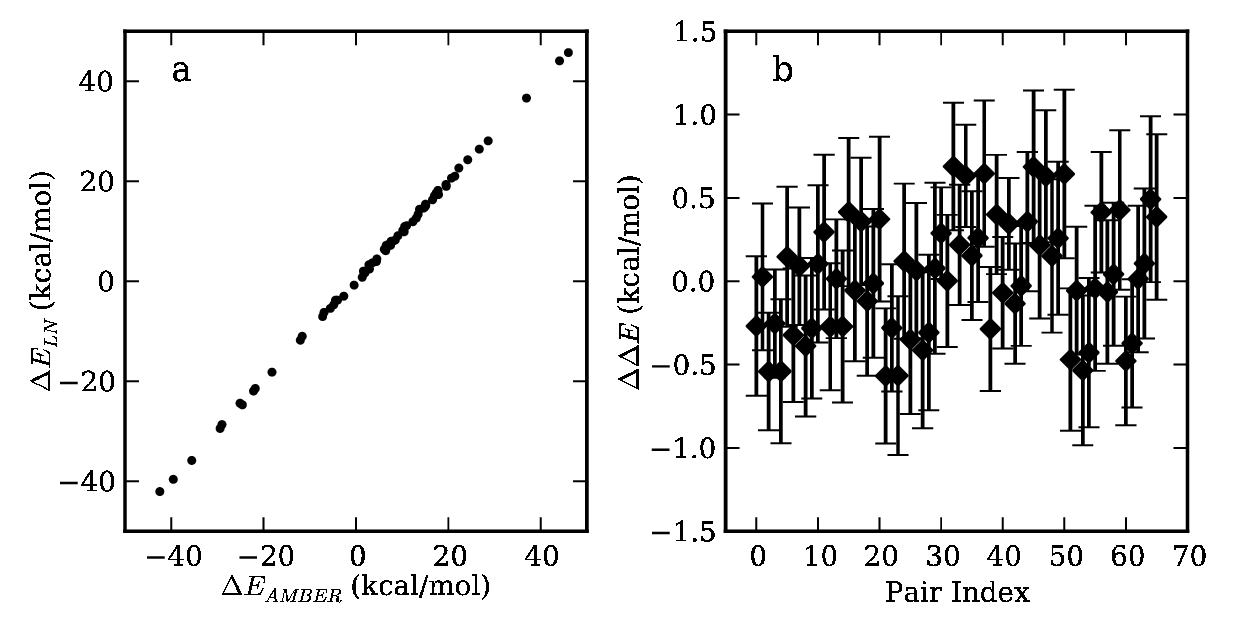}
\caption{
A comparison of AMBER and LN using total internal energy differences between unique pairs in a set of twelve diverse structures of trpzip2.
The energy parameters and simulation conditions were identical to those used for the data in Figure \ref{fig:AMBER_vs_LN_direct}.
(a) Energy differences $\Delta E$ are highly correlated, and (b) the quantity $\Delta \Delta E = \Delta E_{\mathrm{AMBER}} - \Delta E_{\mathrm{LN}}$ is small and distributed about zero.
Moreover, the deviation of $\Delta \Delta E$ about zero decreases with increased sampling (data not shown).
}
\label{fig:AMBER_vs_LN_diff}
\end{figure*}

Figure \ref{fig:AMBER_vs_LN_diff}a shows the differences in average internal energy for each unique pair in the resulting set of structures.
These differences for the two MD algorithms are plotted against one another, showing a close correlation. 
Figure \ref{fig:AMBER_vs_LN_diff}b shows the deviation between LN and AMBER for the energy difference of each pair of structures.
When averaged over all pairs, this deviation is very small, at $0.027 \pm 0.043$ kcal/mol.
Evidently, despite the small systematic errors in absolute energies and temperature seen in the previous section, the LN algorithm preserves energy differences well.

%

\end{document}